\documentstyle[aps,twocolumn]{revtex}
\draft 
\newcommand{\bibi}{\bibitem}                                                  
\newcommand{\etal}{\it {et al.}}                                             

\newcommand{\half}{\frac {1}{2}}                                             
                                             
\newcommand{\beq}{\begin{equation}}                                           
\newcommand{\eeq}{\end{equation}\noindent}                                  
\newcommand{\beqr}{\begin{eqnarray}}                                          
\newcommand{\eeqr}{\end{eqnarray}\noindent}                                   

\newcommand{\vk}{{\bf k}}                                                     
                                                     
\newcommand{\vq}{{\bf q}}

\newcommand{\cd}{c^{\dag}} 
\newcommand{\bd}{b^{\dag}}

\begin{document}     
\twocolumn[\hsize\textwidth\columnwidth\hsize\csname@twocolumnfalse\endcsname
\title{Spin-Charge Recombination and Superconductivity in the $t-J$ Model} 

\author{Sanjoy K. Sarker \\                                                   
Department of Physics,                                        
The University of Alabama, Tuscaloosa, AL 35487 \\                           
and \\
Department of Physics, 
The Ohio State University, Columbus, OH 43210 \\}
\maketitle 
\widetext                                                                  
\begin{abstract}  
A spin-charge binding mechanism is analyzed
using the Schwinger boson representation of the 2-D t-J model. 
In the normal state spinons are paired into singlets which
evolove out of the antiferromagnetic phase at half-filling.
The singlets induce, via the hopping term,
a pairing of holons and bind with the latter. The $d_{x^2-y^2}$ 
symmetry of the order parameter and the shape of the $T_c$ vs doping 
curve, observed in cuprate superconductors, emerge naturally as a 
consequence of the symmetry of the spinon pairs, thus
connecting superconductivity with quantum spin fluctuations in
the insulator.


                               
\end{abstract}
]
\narrowtext
 
The normal state of cuprate superconductors is essentially 
two-dimensional, has a large Fermi surface but apparently hole-like 
charge carriers and shows non-Fermi liquid charateristics. Anderson 
has argued that this is due to spin-charge separation in a doped 
quantum antiferromagnet \cite{and}. In this picture, charge is 
carried by spinless holes of concentraion $\delta$ moving about in 
a quantum liquid of spin-singlets. The electron itself is not a
coherent excitations.  The singlets evolve out of the undoped phase. 
Their presence is consistent with the appearance of a spin-gap in the
normal state \cite{tai}, and may have something to do with 
superconductivity, as many now believe. However, the electron 
excitations in the superconducting state seem to be coherent, appearing 
as sharp peaks in the angle-resolved photemission spectrum \cite {camp}. 
Yet, as recent experiments show \cite{norm}, the \lq\lq quasiparticle" 
peaks do not exist above $T_c$. Hence, if the normal state has 
spin-charge separation, superconductivity must be accompanied by 
spin-charge recombination. 

Based on the 2-D $t$-$J$ model, a spin-charge binding mechanism for
superconductivity was originally proposed in 1992, in which spin
singlets induce a pairing of holons and bind with the latter in
order to recover the kinetic energy lost in forming the magnetic
moments \cite{sar1}.  Because of the pivotal role played by the 
singlets, it is essential that the theory be consistent with the 
behavior at $\delta = 0$. Such a description was developed using the 
Schwinger boson representation. This choice is natural since a mean-field 
approximation gives a good description of Neel ordering and the singlets 
at half-filling ($\delta = 0$) \cite{sar2}. The singlets are described 
in terms of a paired spinon condensate with $p$-wave symmetry, which 
evolves into the doped phase where the holons form a Fermi liquid. 

Here we focus on the superconducting transition and show that the nature of 
the superconducting state is intimately connected with that of the insulator 
through the singlets. We will show that the superconducting
order parameter is {\em constrained} to have the observed $d_{x^2-y^2}$ 
symmetry as a consequence of the ($p$-wave) symmetry of the paired
spinons.  The observed shape of the $T_c$ vs doping curve (showing
a maximum) occurs naturally in our theory, and at small $\delta$, 
$T_c \propto e^{-1/\delta}$, also due to symmetry and statistics. 
The theory predicts the opening of a gap at $T_c$ which has interesting
observable consequences in the underdoped region. 

Our approach is based on the following. (1) Spin-charge recombination 
occurs directly through the hopping term, and in both the normal and 
superconducting states. The energetic reason is as follows. At 
half-filling one has only spins (local moments).  A \lq\lq spin" is 
localized charge, and costs kinetic energy of order $t$, 
which the system tries to recover upon doping. Since kinetic energy gain 
is clearly the largest for free electrons, the best way to recover it is to 
bind spin and charge into electrons and restore the Fermi liquid.
But to move coherently, electrons must go around each other. 
This will happen for large dimensions (e.g., $d \ge 3$),
but clearly not in $d = 1$. The situation is intermediate in 
$d = 2$ \cite{sar1}. (2) Recombination is a {\em collective phenomenon},
and thus is missing at the mean-field (i.e., Hartree-Fock) level. 
It is taken into account by a random-phase approximation (RPA) in which 
the electron appears as a collective excitation. There are some
misconceptions about the Schwinger-boson representation because the Fermi 
surface consists of pockets in the mean-field approximation \cite{sar3}.
By including RPA we have shown that, spin-charge recombination in the 
($d = 2$) normal state is strong enough for electrons to appear deep in the 
Fermi sea, where the kinetic energy gain is large. This is sufficient to 
restore the large Fermi surface and destroy long-range magnetic order. 
However, close to the Fermi surface electrons remain incoherent \cite{sar1}. 
(3) Then the singlets give rise to superconductivity by binding with
holon pairs to recover additional kinetic energy.  This process is taken
into account by anomalous RPA, as shown below.  A similar mechanism, but 
with an on-site interaction, has been recently studied in the $SU$-2 version 
of the slave-boson theory \cite{lee}. However, the slave-boson approach is 
rather different because of different statistics, the additional symmetry 
and a very different normal state since recombination processes described 
above are not considered.

The $t-J$ model is described by the Hamiltonian 
\beq H = - t\sum _{ij,\sigma} \cd _{i\sigma}c_{j\sigma} ~+~ 
 J\sum _{ij} (\vec S_i.\vec S_j ~ - \frac{1}{4} n_in_j), \eeq
subject to the constraint that no two electrons can occupy the 
the same site. Here the sum is over nearest neighbors, and
for cuprates, $t/J \sim 3-4$. The projected electron operator
$\cd_{i\sigma}$ is decomposed as
$\cd _{i,\sigma} = \bd _{i\sigma}h_i$, where $\bd _{i\sigma}$ 
creates a bosonic spinon of spin $\sigma$ and $h_i$ destroys a 
fermionic holon. Then the charge and spin densities can be 
written as $n_i = h^{\dag}_ih_i$, 
$S^{+} = \bd _{i\uparrow}b_{i\downarrow}$ and 
$S^z_i = \half \lbrack \bd _{i\uparrow}b_{i\uparrow} - 
\bd _{i\downarrow}b_{i\downarrow}\rbrack$,
provided that the constraint $h^{\dag} _ih_i + 
\sum _{\sigma}\bd _{i\sigma}b_{i\sigma} =  1$ is satisfied.
We will treat the constraint on the average. However, the
representation has a remarkable property. Even when constraints are 
treated on the average, the condition that electrons (and Cooper pairs) 
avoid each other is always satisfied since
\beq \cd_{i\uparrow}\cd_{i\downarrow} = 0 \eeq
because holons are fermions ($h_ih_i = 0$). Therefore, a crucial 
part of the large-$U$ physics is automatically taken into account in our 
theory.

We will treat the exchange term at the Hartree-Fock level. But
RPA contribution will be included in treating the
hopping term. This term describes spinon-holon interaction:
\beq \sum \epsilon(\vk)\cd_{\vk\sigma}c_{\vk\sigma}
= \frac{1}{N} \sum \epsilon(\vk) 
\bd_{\vk+\vk_1,\sigma}h_{\vk_1}h^{\dag}_{\vk_2}b_{\vk+\vk_2,\sigma}, \eeq
where the potential $\epsilon (\vk) = - 2t(\cos k_x + \cos k_y)$ 
is simply the free-electron hopping energy. As shown earlier \cite{sar3}, 
in the mean-field approximation holons form a cosine band.
The mean-field spinon Hamiltonian is given by
$$ H_b = \sum _{\vk,\sigma} \epsilon _b(\vk) \bd _{\vk,\sigma}b _{\vk\sigma}
-\sum _\vk \phi (\vk) \lbrack \bd _{\vk\uparrow}\bd_{-\vk\downarrow}+
h.c.\rbrack, $$ 
where $\epsilon _b(\vk)$ is associated with hopping. The second term
describes singlet physics through spinon pairing with an
anomalous self energy
\beq \phi (\vk) = 2JA (\sin k_x + \sin k_y). \eeq 
where $A$ is the magnitude of the singlet pairing operator 
$\half \lbrack b_{i\uparrow}b_{j\downarrow} - 
b_{i\downarrow}b_{j\uparrow}\rbrack$. The symmetry of $\phi (\vk)$
is determined at half-filling and remains unchanged in the doped
phase. As far as magnetic properties are concerned, spinon pairing is 
analogous to superconducting pairing, and leads to a spinon pseudogap and 
a suppression of the uniform magnetic susceptibility below the 
pairing temperature $kT_{SG} \sim JA$. Diagonalization leads to a spinon 
spectrum which, near its minimum at $\vk_0$, is approximately given by 
$\omega _b(\vk) = \lbrack \Delta ^2_s + c^2_s(\vk-\vk_0)^2\rbrack ^{1/2}$, 
where $c_s$ is related to the spin-wave velocity, and 
$\Delta _s \equiv \omega _{min} \ge 0$. Bose condensation at $\vk_0$ 
gives rise to magnetic order (LRO) at ${\bf Q} = 2\vk_0$. 
At half filling, ${\bf Q} = (\pi,\pi)$, but is incommensurate away 
for $\delta > 0$. Generally $\Delta _s = 0$ in the condensed phase, 
but becomes finite once LRO is destroyed.

As shown in ref. (5), LRO is destroyed by spin-charge recombination
in the normal state, which is taken into account by RPA. The 
mean-field electron Green's function $G_{c0}$ is the convolution 
(bubble) of spinon and holon Green's functions: $G_{c0}(p) = 
(N\beta)^{-1}\sum _{q} G_b(q)G_h(p+q)$, where $\beta = 1/kT$ and  
$p = (\vk,\omega)$, $\omega$ is the Matsubara frequency. In the 
absence of LRO, $G_{c0}$ is incoherent \cite{sar4}. Summing the 
bubbles, we obtain the
the RPA expression for normal-state Green's function
\beq G_{cn}(p) = \frac{G_{c0}(p)}{1 - \epsilon (\vk)G_{c0}(p)}. \eeq
This is characterized by a large Fermi surface and has poles
for $\vk$ deep in the Fermi sea, but none close to the Fermi surface 
\cite{sar1}. 

Thus the low-energy physics of the normal state is described by a Fermi 
liquid of holons which have with their own Fermi surface 
centered at $(\pi,\pi)$, and a condensate of paired spinons. It is useful 
to note that since $G_{c0}$ is a convolution, a gap in either the 
spinon or the holon spectrum will induce a gap in $G_{c0}$ so that 
$Im  G_{c0} = 0$ below some frequency. This will in turn lead to 
gap in $G_{cn}$ since from  Eq. (5) $Im  G_{cn} =  0$, unless $G_{cn}$
has poles in the gap region. In our case, there are no poles in the
low energy region. 

Superconductivity involves another type of spin-charge binding 
which requires summing anomalous bubbles.  We use functional 
integral method involving three fields: $b_{p\sigma}$, $h_p$
and an auxilliary fermionic field $a_{p\sigma}$ which is 
introduced through a Hubbard-Stratonovitch 
transformation on the hopping term, as described in ref.(5).
The hopping part of the action is then given by:
\beq {\cal A}_{hopp} = \sum _{p\sigma}\epsilon ^{-1}(\vk) 
a^*_{p\sigma}a_{p\sigma} + \lbrack a^*_{p\sigma}c_{p\sigma} + 
c.c.\rbrack, \eeq
where the electron field is the composite object
$c_{p\sigma} = (N\beta)^{-1/2}\sum _qh^*_qb_{p+q}$. Therefore the 
term in the bracket describes the interaction. The remaining
part of the action is quadratic.  The $a$-field is conjugate to 
the $c$-field and their Green's functions 
are simply related.  Let $G_a(p) = <a^*_{p\sigma}a_{p\sigma}>$ 
and $F_{a}(p) = <a_{-p\downarrow}a_{p\uparrow}>$ be the normal and
anomalous Green's function, respectively.  Similarly, for the other 
fields. Then electron Green's function are found from the 
identities: $G_a(p) = \epsilon (\vk) + 
\epsilon ^2(\vk)G_c(p)$, $F_{a}(p) = \epsilon ^2(\vk)F_{c}(p)$. 
Hence the introduction of $a$-fields allows us to treat 
spinons, holons and electrons in a symmetric fashion.  Physically 
$G_a$ is the renormalized hopping potential. Thus, the bare function 
is: $G_{a0} = \epsilon (\vk)$. 

Next we compute the self energies self-consistently to leading order,
which corresponds to Hartree-Fock and RPA.  We use the subscript $n$ 
and $A$ for normal and anomalous functions, respectively. 
Since the three fields appear symmetrically the self-energy of one 
field is simply the bubble of the other two.  Thus the normal 
self-energy $\Sigma _{an}$ is the bubble $G_{c0}$. This gives 
\beq G_{an}(p) =
\frac{\epsilon (\vk)}{1 - \epsilon (\vk)G_{c0}(p)}, \eeq
which is the {\em screened} hopping potential, and
leads to the RPA Green's function (Eq. 5).

Now, since spinons are paired $F_{b}(p) = <b_{-p\downarrow}b_{p\uparrow}>$ 
is nonzero in the normal state. Then, if we integrate out
the spinons we obtain an interaction of the form \cite{sar1} 
\beq \sum a^*_{p\uparrow}a^*_{-p\prime\downarrow}h^*_{q-p}h^*_{-q+p\prime}
F_b(q)~+~c.c. \eeq
Clearly such an interaction encourages pairing. We can
compute the relevant anomalous self-energies (\lq\lq gap functions")
by allowing the holon pairing amplitude 
$F_{h}(p) = <h_{-p}h_p>$ to be nonzero (which ensures that 
$F_a \ne 0$) and doing a pairwise decomposition of Eq. (8).
The result is
\beq \Sigma _{aA}(p) = \frac{1}{N\beta}\sum _{p_1}
F^*_{h}(p_1)F_{b}(p-p_1), \eeq
\beq \Sigma _{hA}(p) = \frac{1}{N\beta}\sum _{p_1}
F^*_{a}(p_1)F_{b}(p_1-p), \eeq
\beq \Sigma _{bA}(p) = \phi (\vk) + \frac{1}{N\beta}\sum _{p_1}
F_{a}(p_1)F_{h}(p-p_1), \eeq
where, for completeness, we have included the additional 
contribution to the $\Sigma _{bA}$ due to holon and electron pairing. 
However $\phi (\vk)$, which is nonzero in the normal state, is the driving 
term, i.e., we seek solutions such that all three $\Sigma _A$ vanish 
if $\phi (\vk) = 0$.  The Green's functions are to be calculated 
self-consistently. They have the same general form
\beq F(p) =  \Sigma _{A}(p)/\lbrack (G^{-1}_{n}(p) G^{-1}_{n}(-p)
 \pm \Sigma ^*_{A}(p)\Sigma _{A}(p)\rbrack, \eeq
where the minus sign in the denominator is for $F_{b}$. The inclusion
of the anomalous bubble $\Sigma _{aA}$ in the Green's functions is
equivalent to summing these bubbles to all orders, thus taking into
account spin-charge recombination in the superconducting state.

Note that, if $\Sigma _A(-p) = \pm \Sigma _A(p)$, then Eq. (12) yields
$F(-p) = \pm F(p)$. (Here $-p = (-\vk,-\omega)$).  Since the holons 
are single-component fermions, $F_h(-p) = - F_h(p)$. Now, the
driving term $\phi _{\vk}$ is odd. Let us therefore assume 
the full $\Sigma _{bA}$ is odd.  Then it follows from Eqs. (9-10) that
$\Sigma _{aA}(p)$ is even, $\Sigma _{hA}(p)$ is odd. Using these in
Eq (11), we find that $\Sigma _{bA}$ is odd, thus completing the
self-consistency loop.

The holon pairing problem is analogous to the BCS problem with
spinons playing the role of phonons. As a first approximation
we specialize to the low-frequency or or weak-coupling (i.e., BCS) limit. 
This will enable us to extract some important results that are
qualitatively correct. First we ignore the second term in Eq.(11)
so that $\Sigma _{bA} \approx \phi (\vk)$.
Next we put $\omega = 0$ in self-energies and define the gap functions
$\Delta _{a,h}(\vk) \equiv \Sigma _{a,hA}(\vk,\omega = 0)$. 
The effective action for the holons can be written as
$${\cal A}^{eff}_h = \sum_p \lbrack i\omega - \epsilon _h(\vk))h^*_ph_p 
+ \frac{\Delta _h(\vk)}{2}(h^*_ph^*_{-p} + h_{-p}h_p)\rbrack,$$ 
where $\epsilon _h(\vk)$ is the effective holon energy. This is
formally identical to the BCS problem. The holons are gapped with energy 
$E_h(\vk) = \lbrack \epsilon^2_h(\vk) + \Delta^2_h(\vk) \rbrack ^{1/2}$.
The difference here is that $\Delta _h$ (roughly $\sim kT_c$) is
determined by holon pairing as well as spin-charge binding. 

At low frequencies, $F_{b}$ can be approximated by its mean-field
form: $F_{b}(\vk,\omega) = \phi (\vk)/\lbrack \omega ^2 + 
\omega ^2_b(\vk)\rbrack$, but with a renormalized spinon energy 
$\omega _b{\vk}$. Since spinons are pseudogapped, 
we put $\omega = 0$ and replace $\omega _b (\vk)$ by a constant
$\omega _0$ which will play the role of a frequency cut-off.
This yields the coupled gap equations
\beq\Delta _h(\vk) = \frac{1}{N\omega ^2_0} 
\sum _{\vq}\phi (\vq-\vk)\Delta _a(\vq)g_a(\vq),\eeq
\beq\Delta _a(\vk) = \frac{1}{N\omega ^2_0} 
\sum _{\vq}\phi (\vk-\vq)\Delta _h(\vq)g_h(\vq),\eeq
where $\beta g_{a,h}(\vk) 
\equiv \sum _\omega F^*_{a,h}(\vk,\omega)/\Delta _{a,h}(\vk)$.

Using $\phi (\vk) = 2JA (\sin k_x + \sin k_y)$ and the fact that
that $\Delta _a(\vk)$ is even and $\Delta _h(\vk)$ is odd, we find
from Eqs. (13-14) that the gap functions must have the form 
\beq \Delta _a(\vk) = a_ 1 \cos k_x + a_2 \cos k_y, \eeq 
\beq \Delta _h(\vk) = h_1 \sin k_x + h_2 \sin k_y, \eeq
where, $a_i$ and $h_i$ are constants.
On general symmetry grounds the magnitude of $\Delta _{a,h}$ 
should be invariant under $k_x \leftrightarrow k_y$.  This implies 
that $\Delta _a(\vk) = a_1(\cos k_x + e^{i\theta _a} \cos k_y)$, 
and $\Delta _h(\vk) = h_1 (\sin k_x + e^{i\theta _h}\sin k_y)$, where 
$\theta _a$ and $\theta _h$ are unknown phases. Finally, using the fact
that in our theory $<\cd_{i\uparrow}\cd _{i\uparrow}> = 0$ is always
satisfied, we have
\beq \sum _{\vk} a_1(\cos k_x + e^{i\theta _a}\cos k_y)g_a(\vk) = 0.\eeq
Now, $g_a(k_x,k_y) = g_a(k_y,k_x)$, which implies $\theta _a = \pi$. 
Then we have
\beq \Delta _a(\vk) = \frac{\Delta _{a0}}{2}(\cos k_x - \cos k_y), \eeq  
where $\Delta _{a0} \equiv 2a_1$.  Noting that the pairing amplitude for $c$
and $a$-fields are proportional, the order parameter for the physical
electron pair has $d$-wave symmetry, solely as a consequence of 
statistics and symmetry of the anomalous spinon self-energy $\phi (\vk)$.
There is thus a deep connection between quantum spin fluctuations in the 
antiferromagnet, the spin-gap phenomenon and the superconducting state. 

Using Eq.(18) in Eq.(13) we find that $\theta _h = \pi$, i.e.,
\beq \Delta _h(\vk) = \frac{\Delta _{h0}}{2}(\sin k_x - \sin k_y), \eeq
where $\Delta _{h0} \equiv 2h_1$. Hence, the holon gap has p-wave 
symmetry.  The gap equations for $\Delta _{a0}$
and $\Delta _{h0}$ are then given by
\beq \Delta _{a0} = - \frac{JA\Delta _{h0}}{N\omega ^2_0}
 \sum _\vk(\sin k_x - \sin k_y)^2g_h(\vk), \eeq
\beq \Delta _{h0} = - \frac{JA\Delta _{a0}}{N\omega ^2_0}
 \sum _\vk(\cos k_x - \cos k_y)^2g_a(\vk). \eeq
The fact that the symmetry is determined in the weak-coupling approximation  
does not limit its generality since by continuity the strong-coupling
function would have the same symmetry. 

We can also obtain an approximate formula for $T_c$ as follows.
At $T_c$ we can put $\Delta _{a,h} = 0$ in the denominator of the
Green's functions  and do the frequency sums \cite{sums}.  Combining 
Eqs. (20) and (21) we get the condition for $T_c$: 
$1 = J^2A^2It^2_R/\omega ^3_0$, where
\beq I =  \int \frac{d^2k}{4\pi ^2} \frac{(\sin k_x - 
\sin k_y)^2}{2\epsilon_h(\vk)} tanh \frac{\epsilon _h(\vk)}{2kT_c}. \eeq
\beq t^2_R = \int \frac{d^2k}{8\pi^2} (\cos k_x - \cos k_y)^2
\epsilon ^2_{scr}(\vk). \eeq
Here $\epsilon _{scr}(\vk) \equiv  G_{an}(\vk,\omega = 0)$ is the
screened hopping energy (Eq. 7). The quantity $t_R$ is a renormalized 
hopping parameter and is independent of $T_c$. 
The integral $I$ describes the physics of holon pairing and,
and apart from the symmetry factors, has the standard (BCS) form.
For small $\delta$ (i.e., in the region of interest) holons form a small
Fermi circle of momentum $k_{Fh} =  (4\pi\delta)^{1/2}$. Then we can 
approximate $\sin k_x$ by $k_x$, etc, and use the
quadratic form for the holon energy: $\epsilon _h(\vk) = 
t_h(k^2 - k^2_{hF})$, where $t_h$ is the renormailized hopping
parameter for holons.  Doing the integral $I$ we obtain
\beq kT_c = \frac{2\omega _1e^{\gamma}}{\pi}e^{-1/\lambda},\eeq
where $\gamma$ is the Euler constant, $\omega _1$ is a frequency
cutoff, and
\beq \lambda = \frac{t^2_RJ^2}{\omega ^3_0t_h}A^2\delta \eeq
is the dimensionless coupling constant that determines $T_c$.
We can take $\omega _1 = min(\omega _0,\epsilon _{Fh})$, where
$\epsilon _{Fh} = 4\pi t_h\delta$ is the holon Fermi energy. A more
elaborate choice does not change the result qualitatively. Based on
previous analysis \cite{sar3,sar1}, we expect the parameters
$t_R$, $t_h$ and $\omega _0$ to vary slowly with $\delta$ compared
with the factor $A^2\delta$ in Eq. (25). The spinon pairing
amplitude $A$ is largest at half-filling, and decreases 
with increasing $\delta$ (already at the mean-field \cite{sar3}) level.
The crucial factor of $\delta$ arises from the symmetry factor 
$(\sin k_x - \sin k_y)^2$ in Eq. (22) which, at the holon Fermi surface,
scales as $\sim (k_x - k_y)^2 \sim k^2_{hF} = 4\pi\delta$. Hence, $T_c$ 
has a maximum, and in particular, 
 $$T_c \propto e^{- const/\delta}$$ 
at small $\delta$, which is a consequence of symmetry and statistics.
Thus the characteristic shape observed in cuprates occurs naturally
in our theory. Furthermore, the exponential dependence on $\lambda$ 
implies that $T_c$ is appreciable only in a small finite region along 
the $\delta$ axis, as observed.  These results are qualitatively
correct, for quantitative accuracy one needs to solve the strong
coupling equations (9-11).

The holons acquire a gap $\Delta _h$ below $T_c$. As discuused earlier
(see below Eq. 5), this leads to a gap in the electron spectrum. Note
that this gap is independent of any pseudogap that may exist in
the normal state. In other words, if the electron already has
a gap  above $T_c$, the total electron gap will show a sharp increase
below $T_c$ as temperature decreases.  Apparently such an increase has 
been seen \cite{norm}.
 
To conclude, we have considered a spin-charge binding mechanism which
differs from traditional approaches in that the electron itself is a 
collective object, and that part of the condensation energy comes from 
hopping. We stress that our results do not change qualitatively if a
nearest-neighbor hopping is included (the $t-t^{\prime}-J$ model) since
symmetry of $\phi (\vk)$ does not change.

The author thanks C. Jayaprakash, W. Puttika, N. Nagaosa, 
T. Lamberger, and particularly T. L. Ho for many discussions.
This work has been supported in part by grants from NSF (DMR 9705295i,
DMR 9807284) and NASA (NAG8-1441).


\end{document}